\documentstyle[twocolumn,pra,aps,epsfig,psfrag]{revtex}
\begin{document}
\flushbottom
\draft
\title{Instabilities and self-oscillations in atomic four-wave mixing}
\author{J.~Heurich, H.~Pu, M.~G.~Moore and P.~Meystre}
\address{Optical Sciences Center,
University of Arizona, Tucson, Arizona 85721
\\ \medskip}

\author{\small\parbox{14.2cm}{\small\hspace*{3mm} The
development of integrated, waveguide-based atom optical devices
requires a thorough understanding of nonlinear matter-wave mixing
processes in confined geometries. This paper analyzes the
stability of counterpropagating two-component Bose-Einstein
condensates in such a geometry. The steady state field equations
of this system are solved analytically, predicting a multivalued
relation between the input and output field intensities. The
spatio-temporal linear stability of these solutions is
investigated numerically, leading to the prediction of a
self-oscillation threshold that can be expressed in terms of a
matter-wave analog of the Fresnel number in optics.
\\
\\[3pt]PACS numbers: 03.75.-b 03.75.Be 03.75.Fi 42.65.Pc}}
\maketitle
\narrowtext
\section{Introduction}
The recent development of narrow atomic waveguides micro-fabricated
on glass chips \cite{1099TWHaenschPRL,1299EACornellPRL,0200MPrentissPRL,0500JSchmiedmayerPRL}
 has raised the exciting possibility of
the design and manufacture of integrated atom-interferometry-based
sensing devices. With the inclusion of an `atom laser'
\cite{0197WKetterlePRL,1198MAKasevichSci,0399WDPhillipsSci,0499TEsslingerPRL}
as a high-brightness source of coherent atomic matter-waves, it is
possible to imagine `practical' devices which could compete with
or out-perform conventional optical interferometric sensors. The
use of high-density atomic fields comes at a price, however, as
atomic matter waves are subject to nonlinear wave mixing due to
atom-atom interactions. It is of crucial importance, therefore, to
understand the effects of nonlinear wave mixing on wave-guide
based atom-optics devices so that they may eventually be
controlled or even exploited. In this spirit, the present paper is
a first attempt at an analysis of wave-mixing instabilities in
quasi-one-dimensional ultracold atomic samples.

The observation of atomic four-wave mixing
\cite{0795PMeystreQSO,0796PMeystreJRN,0299PMeystrePRA,1298PSJulienneOEx,0399WDPhillipsNat,0200PSJulienneQPH,0400YQWangPRA}
and solitons
\cite{1193EMWrightPRL,0894EMWrightPRA,0498PZollerPRL,1299MLewensteinPRL,0100WDPhillipsSci}
in Bose-Einstein condensates of dilute atomic vapors
\cite{0795EACornellSci,1195WKetterlePRL,0895RGHuletPRL} clearly
demonstrate both the significance of nonlinear effects in quantum
degenerate atomic fields, as well the benefits of exploiting the
mathematical analogy between the nonlinear equations describing
self-interacting Schr\"odinger fields and those describing the
propagation of light in nonlinear media. Nonlinear wave-mixing
instabilities have been studied extensively in nonlinear optics,
and many of the techniques and results developed can readily be
adapted to the problem at hand.

Focusing on effective one-dimensional geometries, the question of
stable and unstable steady-state configurations has long been a
topic of optical research. Winful and Marburger
\cite{0480JHMarburgerAPL} first proposed that bistability could
occur in collinear degenerate four-wave mixing and shortly
thereafter Silberberg and Bar-Joseph \cite{0582IBarJosephPRL}
showed that even for the rather simple case of equally polarized
counterpropagating laser beams instabilities and even chaos may
occur in the dynamical behaviour. Multi-branched steady-state
solutions were first derived by Kaplan and Law
\cite{0985CTLawJQE}, however the stability of the steady-state
field configurations was not determined.  Considerable work on the
spatial or temporal stability of such systems was subsequently
carried out by many others
\cite{1183AEKAplanOLe,0485KOtsukaPRL,0286SWabnitzPRL,0487JESipePRA,0687PWMilonniPRL,0592DNChristodoulidesPRA}.
In particular, optical instabilities and polarization bistability
were experimentally observed in sodium vapor
\cite{1088RWBoydPRL,0490RWBoydPRL}.

\begin{figure} [tf]
  \begin{center}
    \epsfig{file=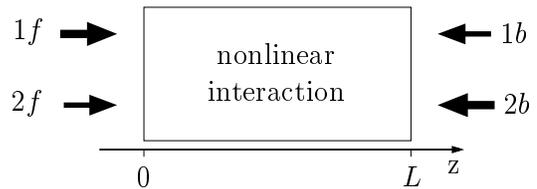, width=0.8\columnwidth}
  \end{center}
\caption[setup] {\label{setup} Four matter waves incident into a
region of nonlinear interaction. The two forward moving and the
two backward propagating modes, with opposite wave vectors, are
distinguished by their internal state.}
\end{figure}

Bistability and nonlinear instability are typically related to
four-wave-mixing phenomena in systems which exhibit a cubic
nonlinearity. In ultracold atomic systems it is readily shown that
in the $s$-wave scattering approximation the form of the
self-interaction is that of a cubic nonlinearity. A recent paper
by Law and coworkers \cite{1298NPBigelowPRL} analyzed
four-wave-mixing processes between the hyperfine ground-state
components of components of a $^{23}$Na spinor condensate confined
in an optical dipole trap. Goldstein and Meystre
\cite{0599PMeystrePRA} presented a full quantum-mechanical theory
of four-wave mixing in a system where two $m_F=0$ momentum side
modes were counterpropagating while the $m_F=\pm1$ states were at
rest.

In the present work we investigate a collinear four-wave mixing
geometry as sketched in Fig.~\ref{setup}, where each of the
counterpropagating matter waves can be in one of two different
atomic states, for example two hyperfine state levels of
$^{87}$Rb. This situation is closely analogous to the optical
case, the two internal atomic states taking the place of the
polarizations of the field. As such, this system is formally
equivalent to the case of counterpropagating light fields in an
 Kerr medium.

In contrast to the exact quantum treatment of
Refs.~\cite{1298NPBigelowPRL,0599PMeystrePRA}, our analysis is
based on a mean-field approach, the matter-wave equivalent of
treating the electromagnetic field classically. We investigate
both the steady-state and dynamical behavior of the system by a
combination of analytical and numerical methods. The output fields
are found to generally exhibit a multi-valued dependence on the
inputs, characteristic of bistable and multi-stable systems. The
stability analysis for this particular configuration, however,
shows that only the upper branch of the steady-state curve is
stable against small perturbations. More interesting, however, is
the occurrence of a threshold behavior in the output fields
indicating the onset of self-oscillations in the system. The
feedback mechanism leading to this effect is the grating
established in the medium by the interference between the various
fields.

We note that matter-wave bistability was recently predicted in a
simple model of a driven nonlinear Gross-Pitaevskii equation
\cite{0100ZBChenMPL} which neglected however the effects of
collisions between the strong driving field and the condensate. In
contrast, the present system includes both the effect of two-body
collisions, and fully accounts for propagation effects.

This paper is organized as follows: Section \ref{model} introduces
our model and derives the nonlinear partial differential equations
describing the propagation of the interacting atomic beams.
Section \ref{sss} solves these equations analytically in steady
state and shows the appearance of multistable solutions and
threshold behavior. The stability of the steady state solutions is
investigated in Section \ref{stab_anal} using numerical methods,
while section \ref{self_osc} discusses the onset of
self-oscillations. Finally, Section \ref{conclusion} is a summary
and conclusion.

\section{Model}
\label{model}

We consider an ultracold two-component Schr\"odinger field ${\hat
{\mathbf \Psi}}({\bf r},t)= ({\hat \Psi}_1({\bf r},t),{\hat
\Psi}_2({\bf r},t))^T$, the indices 1 and 2 labeling the internal
state of the atoms of mass $m$, e.g. two hyperfine ground states.
These fields consist of two counterpropagating plane waves which
interact propagating along the $z$-axis in an interaction region
$0 \le z \le L$, the nonlinear interaction outside this region
being turned off e.g. by tuning a magnetic field to a Feshbach
resonance.

Our starting point for the description of this system is the
many-body Hamiltonian describing the evolution of a two-component
condensate, the effects of collisions being described in the
$s$-wave scattering approximation,
\begin{eqnarray}
    {\cal H}&=&\sum_{j=1,2}\int d^3r {\hat \Psi}^\dagger_j({\bf r},t)
    \left( \frac {\hat{p}^2}{2m} + V_{j} \right) {\hat \Psi}_j({\bf r},t)
    \nonumber \\
    &&+\frac{\hbar}{2}\int d^3r \left( g_1 {\hat \Psi}_1^\dagger
    {\hat \Psi}_1^\dagger {\hat \Psi}_1 {\hat \Psi}_1
    + g_2 {\hat \Psi}_2^\dagger {\hat \Psi}_2^\dagger {\hat \Psi}_2
    {\hat \Psi}_2 \right. \nonumber \\
    &&+ \left. 2g_x {\hat \Psi}_1^\dagger {\hat \Psi}_2^\dagger {\hat
\Psi}_1 {\hat \Psi}_2
\right)
\end{eqnarray}
where the scattering strengths $g_i$ are related to their
respective s-wave scattering lengths $a_i$ by
\begin{equation}
g_i=\frac{4\pi\hbar a_i}{m}.  \label{g}
\end{equation}

For $T \rightarrow 0$, the condensate is well described by a
two-component Hartree condensate wave function $\phi({\bf r},t) =
\left(\phi_1({\bf r},t), \phi_2({\bf r},t)\right)^T$, governed by
the coupled Gross-Pitaevskii nonlinear Schr\"odinger equations
\begin{eqnarray}
i\hbar\dot{\Phi}_1 & = & \left(\frac{\hbar^2}{2m}\nabla^2 +
V_{1}\right)\Phi_1 \nonumber \\
& & + \hbar N
\left[g_s|\Phi_1|^2 + g_x|\phi_2|^2 \right]\Phi_1,
\end{eqnarray}
and similarly for $\Phi_2$ with $1 \leftrightarrow 2$, where the
total atomic density is normalized to $1$ by factoring out the
number of atoms $N$. Quantum fluctuations about this solution can
be analyzed by introducing the mean-field approximation
\begin{equation}
     {\hat \Psi}_i({\bf r},t) \simeq \Phi_i({\bf r},t) + { \delta
\hat{\psi}}_i({\bf r},t),
\end{equation}
where the bosonic operator $ {\delta \hat{\psi}}_i({\bf r},t)$
describes small fluctuations about the mean field $\Phi_i({\bf
r},t) = \langle {\hat \Psi}_i({\bf r},t)\rangle$. This analysis
will be the subject of future work.

We assume that the atomic fields are tightly confined in the transverse
dimension but free to move in the third one such that
the motional degrees of freedom in the $x-y$ plane
are frozen, a situation that
could be realized in atomic waveguides. In that case,
we may factorize the ground state Hartree wave function into a
parallel and a transverse part as
\begin{equation}
\Phi_j({\bf r}, t)=\phi^{(j)}_\bot(x,y)\phi_j(z,t)e^{-i\omega_j t}
\end{equation}
where $\phi^{(j)}_\bot(x,y)$ is taken to be the normalized ground
state of the transverse potential $V_j(x,y)$ with energy $\hbar
\omega_j$. The problem is then reduced to an effective
one-dimensional geometry, with the longitudinal condensate wave
function satisfying the one-dimensional nonlinear Schr\"odinger
equation
\begin{eqnarray}
&&i \hbar
\dot{\phi}_1(z,t)=-\frac{\hbar^2}{2m}\,\frac{\partial^2}{\partial
z^2}\,\phi_1(z,t) \nonumber \\ &&+  \hbar N \left[\eta_1
g_1|\phi_1(z,t)|^2 + \eta_x g_x |\phi_2(z,t)|^2\right]\phi_1(z,t)
\label{nlse}
\end{eqnarray}
and similarly for $\phi_2(z)$ with $1 \leftrightarrow 2$. Here
\begin{eqnarray}
\eta_j &=& \int dx\,dy\,\left|\phi_\bot^{(j)}\right|^4,\quad j=1,2
\nonumber \\ \label{eta} \eta_x &=& \int
dx\,dy\,\left|\phi_\bot^{(1)}\right|^2
\left|\phi_\bot^{(2)}\right|^2.
\end{eqnarray}
We consider the situation where two counterpropagating beams of
matter waves are moving along the axis of the waveguide, with
\begin{equation}
\phi_j(z,t)=\left[\phi_{jf}(z,t)e^{ikz}+\phi_{jb}(z,t)e^{-ikz}\right]e^{-i
\omega t}
\label{2waves}
\end{equation}
where $j= 1,2$ and $\hbar k^2/2m = \omega$. We assume that the
spatial envelopes of these beams vary slowly over a de Broglie
wavelength, the atom optics version of the slowly varying envelope
approximation,
\begin{equation}
\left|\frac{\partial^2}{\partial
z^2}\phi_m\right|\ll k\left|\frac{\partial}{\partial
z}\phi_m\right|\ll k^2\left|\phi_m\right|.
\label{svea}
\end{equation}
With Eqs.~(\ref{2waves}) and (\ref{svea}), Eq.~(\ref{nlse}) yields
\begin{eqnarray}
&&i\left [\frac{\partial}{\partial t} + \left (\frac{\hbar k}{ m}
\right )\frac{\partial }{\partial z} \right ]\phi_{1f} =g_1
\left(|\phi_{1f}|^2 + 2|\phi_{1b}|^2\right)\phi_{1f} \nonumber \\
&&+ g_x \left[\left(|\phi_{2f}|^2+|\phi_{2b}|^2\right)\phi_{1f}
+\phi_{2f}\phi_{2b}^\star\phi_{1b} \right]
\label{motion1}
\end{eqnarray}
and
\begin{eqnarray}
&&i\left [\frac{\partial}{\partial t} - \left (\frac{\hbar k}{ m}
\right )\frac{\partial }{\partial z}\right ]\phi_{1b} =g_2
\left(|\phi_{1b}|^2 + 2|\phi_{1f}|^2\right)\phi_{1b} \nonumber \\
&&+ g_x \left[\left(|\phi_{2b}|^2+|\phi_{2f}|^2\right)\phi_{1b}
+\phi_{2b}\phi_{2f}^\star\phi_{1f} \right] \label{motion2}
\end{eqnarray}
as well as two additional equations with $1 \leftrightarrow 2$. In
these equations we have scaled the nonlinear coupling constants to
the overlap integral $\eta$ and the total particle number $N$ as
\begin{eqnarray}
    g_j & \rightarrow & \eta_j N g_j,\;\;\;j=1,2 \nonumber \\
    g_x & \rightarrow & \eta_x N g_x.
\label{gscale}
\end{eqnarray}
The factors of 2 appearing in Eqs.~(\ref{motion1}) and
(\ref{motion2}) result from the nonlinear nonreciprocity familiar
in nonlinear optics.

\section{Steady state}
\label{sss}

To find the steady state of the system, we proceed by first
observing that the total atomic density
\begin{equation}
\label{totdensity}
\varrho = |\phi_{1f}|^2 + |\phi_{1b}|^2 +
|\phi_{2f}|^2 + |\phi_{2b}|^2
\end{equation}
is a constant of motion. For simplicity we take $g_1=g_2=g_x\equiv g$
(this can be achieved by tuning the scattering lengths
via e.g., Feshbach resonance and/or by adjusting the transverse
potential $V_j$); in the optical analog this corresponds to a
purely electrostrictive Kerr medium.
Dropping the time derivatives from Eqs.~(\ref{motion1}) and (\ref{motion2}) and
introducing the scaled velocity \begin{equation}
\label{vdef}
v=\frac{\hbar k}{mg}
\end{equation}
gives then
\begin{eqnarray}
\label{motion3}
    iv\frac{d \phi_{1f}}{dz} &=& \varrho \phi_{1f} + R \phi_{1b}
    \nonumber \\
    -iv\frac{d\phi_{1b}}{dz} &=& \varrho \phi_{1b} + R^\star
    \phi_{1f},
\end{eqnarray}
and
\begin{eqnarray}
\label{motion3-1}
    iv\frac{d \phi_{2f}}{dz} &=& \varrho \phi_{2f} + R \phi_{2b}
    \nonumber \\
    -iv\frac{d\phi_{2b}}{dz} &=& \varrho \phi_{2b} + R^\star
    \phi_{2f},
\end{eqnarray}
where the introduction of the new variable
\begin{equation}
  R(z) \equiv \phi_{1f}\phi_{1b}^\star + \phi_{2f}\phi_{2b}^\star
\end{equation}
allows us to decouple the evolution of the two internal components
of the field two condensate species $1$ and $2$. This can be
seen from the observation that
\begin{eqnarray}
    \label{motionR}
    \frac{ dR}{dz} & = & \frac{d\phi_{1f}}{dz} \phi_{1b}^\star +
    \phi_{1f}\frac{d\phi_{1b}^\star}{dz}
    + \frac{d\phi_{2f}}{dz}\phi_{2b}^\star +
    \phi_{2f}\frac{d\phi_{2b}^\star}{dz} \nonumber \\
    & = & - \frac{i}{v} 3 \varrho R
\end{eqnarray}
which yields
\begin{eqnarray}
\label{R_0}
R(z) = R_0 e^{- \frac{i}{v} 3 \varrho z}.
\end{eqnarray}
We can therefore determine the steady state for the two
internal states of the condensate separately.

Substituting the new rotating field amplitudes
\begin{eqnarray}
\label{rot}
    \psi_{1f}&=&\phi_{1f} e^{\frac{i}{v}\varrho z}\nonumber \\
    \psi_{1b}&=&\phi_{1b} e^{-\frac{i}{v}\varrho z},
\end{eqnarray}
into Eqs.~(\ref{motion3}) yields the second-order differential
equation
\begin{equation}
\label{motion-final} \frac{d^2 \psi_{1f}}{dz} =
-\frac{i\varrho}{v} \frac{d\psi_{1f}}{dz} + \frac{|R_0|^2}{v^2}
\psi_{1f},
\end{equation}
which shows that the propagation of $\psi_{1f}$ is characterized
by the spatial frequencies
\begin{equation}
  k_\pm = -\frac{\varrho}{2v}\pm\frac{1}{2v}\sqrt{\varrho^2 - 4|R_0|^2}
\end{equation}

It can be shown that
\begin{equation}
M^2 \equiv \varrho^2 -4|R_0|^2 \ge 0
\end{equation}
so that
$\psi_{1f}(z)$:
\begin{equation}
    \psi_{1f}(z) = e^{-\frac{i\varrho}{2v}z}\left[A_{1f}\sin
\left(\frac{M}{2v}z\right)
    + B_{1f}\cos \left(\frac{M}{2v}z \right)\right],
\end{equation}
the constants $A_{1f}$ and $B_{1f}$ being determined by the
boundary conditions. We observe that $A_{1f}$ depends on both
$\psi_{1b}(0)$ and $R_0$, and hence on the boundary conditions
for all four fields,  $\phi_{1f}(0)$, $\phi_{1b}(L)$,
$\phi_{2f}(0)$ and $\phi_{2b}(L)$. Therefore the equations of motion of
all four fields need to be solved and used to calculate the
respective coefficients for any one field. The explicit forms of
the coefficients $A_{\mu i}$ and $B_{\mu i}$, where $\mu =1,2$ and
$i=f,b$ are given in appendix \ref{appb}.

Further analytical progress can be achieved by decomposing the
field $\psi_{1f}$ into a real amplitude and phase following reference \cite{0985CTLawJQE},
\begin{equation}
  \psi_{1f}=\sqrt{\rho_{1f}}\exp ({i\vartheta_{1f}}),
\end{equation}
and concentrating on the field amplitudes only. One finds readily
\begin{eqnarray}
\label{rhomotion}
  \rho_{1f}(z) & = &\beta \pm \sqrt{|\alpha|^2 -
\left(\rho_{1f}(0)-\beta\right)^2}
  \sin\left(\frac{M}{v}z\right) \nonumber \\
  & & + \left(\rho_{1f}(0)-\beta\right)\cos\left(\frac{M}{v}z\right)
\end{eqnarray}
where
\begin{eqnarray}
  \alpha & = &  \frac{1}{2}\left(A_{1f}^2 + B_{1f}^2\right) \nonumber \\
  \beta & = & \frac{1}{2}\left(|A_{1f}|^2 + |B_{1f}|^2\right) ,
\end{eqnarray}
and the sign in front of the square root in Eq.~(\ref{rhomotion})
is determined by the sign of $A_{1f}B_{1f}^\star + A_{1f}^\star B_{1f}$. Similar
relations hold for the other field components.

One advantage of concentrating on the amplitudes $\rho_{\mu i}$
only is that it is sufficient to know one of them to determine the
others. This follows from the fact that
\begin{equation}
\frac{d\rho_{1f}}{dz}=\frac{d\rho_{1b}}{dz}=-\frac{d\rho_{2f}}{dz}=-\frac{d\rho_{2b}}{dz}.
\end{equation}
which allows us to introduce the three conserved quantities
\begin{eqnarray}
    \rho_f &\equiv& \rho_{1f}(z)+\rho_{2f}(z) \nonumber \\
    \rho_b &\equiv& \rho_{1b}(z)+\rho_{2b}(z) \nonumber \\
    \rho_x &\equiv& \rho_{1f}(z)+\rho_{2b}(z) ,
\end{eqnarray}
so that
\begin{eqnarray}
    \rho_{1b}(z) & = & \rho_b - \rho_x + \rho_{1f}(z)\nonumber \\
    \rho_{2f}(z) & = & \rho_f - \rho_{1f}(z)\nonumber \\
    \rho_{2b}(z) & = & \rho_x - \rho_{1f}(z).
\label{cons_laws}
\end{eqnarray}
The existence of these conservation laws was previously pointed
out in the context on nonlinear optics in Refs.
\cite{0985CTLawJQE} and \cite{0487JESipePRA}.

For concreteness, we now consider the specific example where the
intensities of the forward and backward propagating fields are
equal,
\begin{equation}
\label{constraints}
    \rho_f = \rho_b,
\end{equation}
and
\begin{eqnarray}
    \rho_{1b}(L) & = & \rho_b = \varrho/2 \nonumber \\
    \rho_{2b}(L) & = & 0.
\end{eqnarray}
Under these conditions, one finds readily that
\begin{equation}
\label{R0}
    |R_0|^2 = \rho_b\rho_{1f}(L) = \frac{1}{2}\varrho\rho_{1f}(L)
\end{equation}
and
\begin{equation}
\label{alphabeta}
    |\alpha| = \beta = \rho_{1f}(L)/2.
\end{equation}
With Eqs.~(\ref{R0}) and (\ref{alphabeta}), Eq.~(\ref{rhomotion})
reduces to the remarkably simple form
\begin{equation}
\label{bistab}
\frac{\rho_{1f}(0)}{\rho_{1f}(L)}  =
\cos^2\left(\kappa L/2 \right)
\end{equation}
where
\begin{equation}
    \kappa=M/v
    =\left(\frac{2mg}{\hbar k}\right)
    \varrho\sqrt{1/4 -\rho_{1f}(L)/2\varrho}.
\label{kappa}
\end{equation}
$1/ \kappa$ defines a characteristic length. Equation (\ref{bistab}) predicts a
multivalued relationship between the input and output intensities
$\rho_{1f}$. The longer $\kappa$, the higher order of
``multistability'' predicted by Eq.~(\ref{bistab}). Fig.~2 illustrates
the input-output relationship of $\rho_{1f}$.

Alternatively, and recalling Eqs.~(\ref{gscale}) and (\ref{vdef})
we observe that the argument $\kappa L /2$ of the cosine function
in Eq.~(\ref{bistab}) is also proportional to the total density of
atoms. This leads us to expect some interesting behaviour when
varying the number of particles involved, a scheme whose optical
analog has been the object of considerable work
\cite{0985CTLawJQE,0485KOtsukaPRL,0286SWabnitzPRL,0487JESipePRA}.
However, we don't further investigate these questions in the
present paper, which is limited to the case of fixed total atomic
density.

\begin{figure}[tf]
  \begin{center}
    \epsfig{file=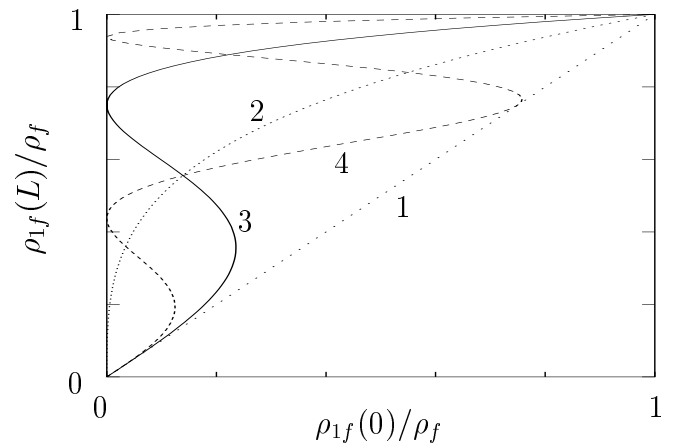, width=1\columnwidth}
  \end{center}
\caption[num1]{\label{bicurve}Normalized output intensity
$\rho_{1f}(L)$ versus normalized input intensity $\rho_{1f}(0)$
from Eq.~(\ref{bistab}). 1:~$L/L_c= 0$, 2:~$L/L_c =2 \pi$,
3:~$L/L_c = 4\pi$ and 4:~$L/L_c = 8\pi$, with $L_c$ as defined in (\ref{Lc}).}
\end{figure}

Returning from the scaled version (\ref{gscale}) of the coupling
constant $g$ to its original definition (\ref{g}) via
Eqs.~(\ref{eta}), we observe that for a transversely homogeneous
sample, the factor $g\varrho$ in Eq.~(\ref{kappa}) becomes $g
\varrho_V$, where $\varrho_V$ is the volumetric density of the
atomic system (as opposed to its linear density $\varrho$).
Eq.~(\ref{kappa}) becomes then
\begin{eqnarray}
\kappa &=& k\left ( \frac{2m}{\hbar^2 k^2} \right )\hbar g \varrho_V
\sqrt{1/4 -2\rho_{1f}(L)/2\varrho} \nonumber \\
&=& k \left(\frac{E_{mf}}{E_{ke}}\right )\sqrt{1/4 -\rho_{1f}(L)/2\varrho}.
\label{kappaf}
\end{eqnarray}
where we have introduced the kinetic energy $E_{ke} =
\hbar^2k^2/2m$ and the mean-field energy $E_{mf} = \hbar g
\varrho_V$.

Introducing the healing length
\begin{equation}
\ell_h = \sqrt{\frac{\hbar}{2mg\varrho_V}}
\end{equation}
and the de Broglie wavelength $\lambda_{db} = 2\pi /k$ shows that
the multistable properties of the system are fully determined by
the characteristic length
\begin{equation}
\label{Lc} L_c = \frac{E_{ke}}{kE_{mf}} = 4\pi^2
\frac{\ell_h^2}{\lambda_{db}}.
\end{equation}

\section{Stability analysis}
\label{stab_anal}

In this section, we analyze the stability of the steady-state
solution (\ref{bistab}) against small classical perturbations. For
simplicity, we assume that the fields at the boundaries of the
interaction region are real,
\begin{eqnarray}
    \phi_{1f}(0) &=& \sqrt{\rho_{1f}}\nonumber \\
    \phi_{2f}(0) &=& \sqrt{\rho_f - \rho_{1f}}\nonumber \\
    \phi_{1b}(L) &=& 0 \nonumber \\
    \phi_{2b}(L) &=& \sqrt{\rho_b}.
\end{eqnarray}

The first approach proceeds by expressing the condensate wave
function $\phi_{1f}(z,t)$ and its complex conjugate as
\begin{eqnarray}
    \phi_{1f}(z,t) & = & \phi_{{1f}_{\displaystyle{ss}}}(z,t) +
\delta_{1f}(z)e^{-i
    \lambda t}\nonumber \\
    \phi_{1f}^\star(z,t) & = &
    \phi_{{1f}_{\displaystyle{ss}}}^\star(z,t) + \epsilon_{1f}(z)e^{-i
\lambda t}
\end{eqnarray}
and linearizing the equations of motion (\ref{motion1}) and
(\ref{motion2}), and similarly for the other field components. The
spectrum of eigenvalues $\lambda$ can be found by discretizing the
resulting system of equations with $N$ points along the $z$-axis
which yields a sparse $8N \times 8N$ eigenvalue problem that can
be solved numerically using standard techniques. The steady-state
solution is then unstable only if eigenvalues with positive
imaginary parts appear in the spectrum.

In addition to this temporal stability, we have also carried out a
full spatio-temporal analysis by solving directly the linearized
form of Eqs.~(\ref{motion1}) and (\ref{motion2}) for small
perturbations about the steady state of the fields.

\begin{figure} [tf]
  \begin{center}
    \epsfig{file=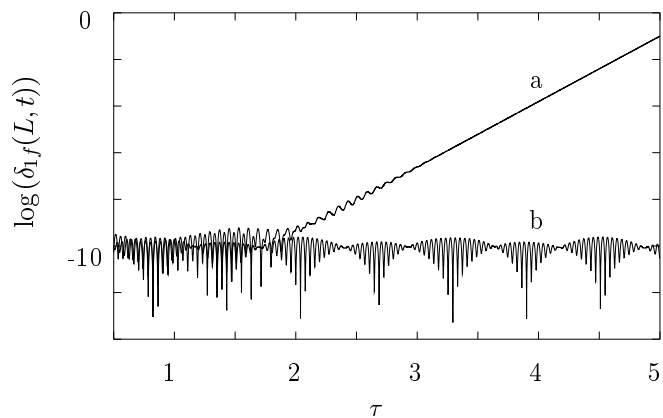, width=1\columnwidth}
  \end{center}
\caption[pert]{\label{pert} $\log\left(\delta_{1f}(L,\tau)\right)$
$\rho_{1f_{\displaystyle ss}}(L)=0.2\rho_f$. $\delta_{1f}(L,\tau)$
is scaled to the intensity $\rho_f$ of the forward moving field
and the time $\tau$ is in units of the round-trip time through the
interaction region. Curve (a) is an example of unstable behaviour
at $L/L_c=4\pi$ ($\kappa L/2 \approx 0.89 \pi$), while curve (b)
shows stability at $L/L_c=1.6\pi$ ($\kappa L/2 \approx 0.36
\pi$).}
\end{figure}

Both approaches indicate that the only stable branch of the
solution (\ref{bistab}) is the ``uppermost branch'', i.~e.~the branch
corresponding to the highest $\rho_{1f}(L)$ for a given
$\rho_{1f}(0)$ which
corresponds to the condition
\begin{equation}
    0\le \frac{\kappa L}{2}< \pi/2 .
    \label{stab_cond}
\end{equation}

Figure \ref{pert} shows a typical result for the temporal
evolution of $\delta_{1f}$ at $z=L$ in both an unstable and a
stable branch. After some short-time transients, the unstable
dynamics becomes completely dominated by the largest positive
eigenvalue and the growth of the small perturbation becomes
exponential.

\section{Self-oscillations}
\label{self_osc}

\begin{figure}[tf]
  \begin{center}
    \epsfig{file=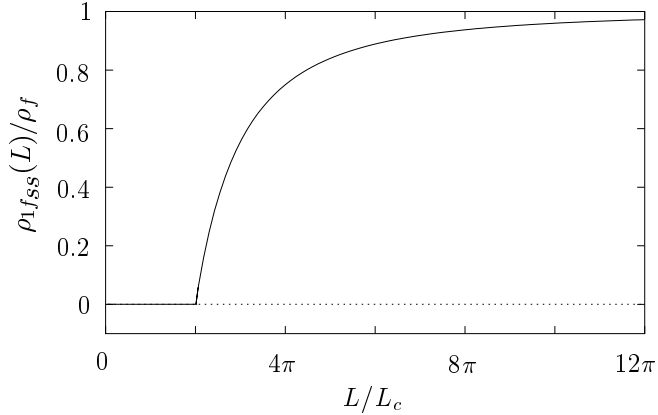, width=\columnwidth}
  \end{center}
\caption[instab]{\label{instab}Self-oscillation threshold of the
intensity $\rho_{1f}(L)$ as a function of the \ interaction length
$L$. Above threshold, this field spontaneously builds up from
fluctuations even in the absence of input, $\rho_{1f}(0) = 0$.}
\end{figure}

An important consequence of the stability analysis is the
prediction of a self-oscillation threshold in the system, as can
be seen by considering the case $\rho_{1f}(0)=0$, that is, no
input field in mode ``$1f$''. In that case, Eq.~(\ref{bistab})
reduces to
\begin{equation}
\rho_{1f}(L)\cos^2(\kappa L/2)=0.
\end{equation}
Hence in the stable region given by (\ref{stab_cond}), $\rho_{1f}(L)$
must be 0; while in the unstable region, $\rho_{1f}(L)$ may take
finite values which results from the amplification of fluctuations.
The onset of instability is given by
\begin{equation}
\kappa L = kL \left ( \frac{E_{mf}}{E_{ke}}\right ) \sqrt{1/4-
\rho_{1f}(L)/2 \varrho} = \pi.
\label{threee}
\end{equation}
Since the argument of the square root is always less than 1/4, we have
that
\begin{equation}
    \kappa L \le \frac{kL}{2} \left ( \frac{E_{mf}}{E_{ke}}\right ).
\end{equation}
Hence the threshold condition (\ref{threee}) cannot be met unless
\begin{equation}
    \left ( \frac{E_{mf}}{E_{ke}}\right ) \ge \frac{2\pi}{kL} ,
\end{equation}
or
\begin{equation}
\frac{\ell_h^2}{L \lambda_{db}} \le \frac{1}{8 \pi^3}
\label{threshold}
\end{equation}
which can be recast as
\begin{equation}
\frac{L}{L_c} \ge 2\pi.
\end{equation}
If this threshold condition is satisfied, the field $\phi_{1f}$
undergoes a second-order-like phase transition to
self-oscillations, as illustrated in Fig. \ref{instab}.

The physical origin of the self-oscillations is the distributed
feedback resulting from the cross-phase modulation with the other
fields present. In order to lead to gain at the wavelength of
$\phi_{1f}$, this grating must itself be modulated with period
$2\pi/k$. But the mean-field energy of the condensate fights
against the creation of such spatial modulation. The healing
length, whose associated
momentum $\ell_h^{-1}$ yields a kinetic energy contribution equal
to the mean-field energy, is the smallest length scale over which
the required changes can occur. Hence, it is not surprising that
the threshold condition should be related to the healing length, the
length of the sample and the atomic de Broglie wavelength.

It is interesting to note that the quantity $\ell_h^2/
\lambda_{db} L$ is reminiscent of the Fresnel number ${\cal F} =
a^2/\lambda L$ in optics, where $a$ is the aperture of the system,
$\lambda$ the wavelength and $L$ the distance of propagation. The
Fresnel number is a measure of the number of transverse modes that
can be excited in an optical system. For very large Fresnel
numbers, the wave can well be approximated as a plane wave, while
diffraction effects and multiple transverse modes become important
for small ${\cal F}$. The present situation is different in that
we now consider the longitudinal stability of the system. Still,
the analogy is rather telling. For samples of length short or
comparable to the healing length, the condensate is so stiff as to
prevent the generation of higher longitudinal modes, hence the
instability requires a condensate much larger than $\ell_h$. As
such, $\ell_h^2/\lambda_{db}L$ can be thought of as the {\em
longitudinal Fresnel number} of the condensate, with the healing length
playing a role similar to that of the system aperture $a$ in
optics.

\begin{figure} [tf]
  \begin{center}
    \epsfig{file=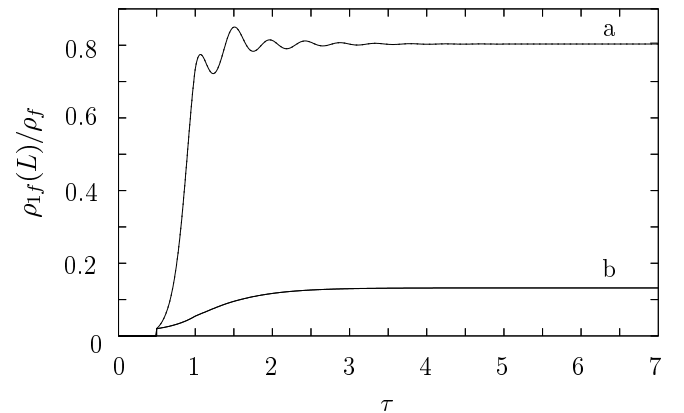, width=\columnwidth}
  \end{center}
\caption[evolve]{\label{evolve} $\rho_{1f}(L,\tau)$ for
$\rho_{1f}(0)=0.01 \varrho$. The time is in units of the
round-trip time through the interaction medium. Curve (a) is in
the regime with only one high output stable branch at
$L/L_c=4\pi$; curve (b) is at $L/L_c=1.6\pi$ where for every (low)
output intensity there exists a stable steady-state field
configuration.}
\end{figure}

The onset of self-oscillations is illustrated in Fig.
\ref{evolve}, which shows the evolution of $\rho_{1f}(L,t)$ for
$\rho_{1f}(0,t)= 0.01 \varrho/2$, this small value simulating some
small fluctuation about $\rho_{1f}(0,t)=0$. This simulation
assume that all field amplitudes inside the interaction region
($0<z \le L$) are
initially equal to zero.

\section{Summary and conclusion}
\label{conclusion} In summary, we have studied a system of
counterpropagating two-component Bose-Einstein condensates in a
waveguide configuration. This system exhibits interesting
nonlinear behavior such as four-wave mixing and self-oscillations.
We have presented an analytical solution of the steady state field
equations and found a multivalued relationship between the input
and the output field intensities. Through a linear perturbation
analysis as well as a full numerical study, we have found that one
and only one branch of solutions is stable for a given input
configuration. The onset of the instability depends on a parameter
in close analogy with the Fresnel number in optics.

We remark that these results were obtained under the assumption
that all the effective scattering lengths are the same
($g_1=g_2=g_x$). In future work, we plan to investigate the case
of unequal scattering lengths, which will make our system formally
equivalent to the case of counterpropagating light fields in a
non-electrostrictive Kerr medium. The analogy with the optical
system also suggests that it would be desirable to allow for
different intensities of the two counterpropagating matter-wave
fields. It has been shown that bi- or multistable solutions exist
in these nonlinear optical systems. It will be interesting to see
if such a matter wave system can also exhibit multi-stability.

Finally, it will be worthwhile to extend the analysis past the
Hartree mean-field theory in order to study the onset of
nonclassical effects, such as the squeezing associated with the
anomalous density of the system.

The experimental realization of this system is certainly
challenging, but we see no fundamental problems using present day
technology. Indeed, the question of stability in wave-mixing
processes will most likely arise naturally in the next generation
of atom interferometry experiments.

\acknowledgments
This work is supported in part by Office of Naval Research Contract No.~14-91-J1205,
National Science Foundation Grant PHY98-01099, the Army Research Office and the
Joint Services Optics Program.
J. H. gratefully acknowledges support by the ``Konrad Adenauer
Stiftung''.

\appendix
\section{}
\label{appb} Solving Eq.~(\ref{motion-final}) and its counterpart
for the backward moving field and inserting the results into Eqs.
(\ref{rot}) gives
\begin{eqnarray}
\label{phimot} \phi_{1f}(z) & = &
    e^{-i\frac{3\varrho}{2v}z}\left[A_{1f}
    \sin\left(\frac{M}{2v}z\right) + B_{1f}
    \cos\left(\frac{M}{2v}z\right)\right]\nonumber \\
    \phi_{1b}(z) & = & e^{i\frac{3\varrho}{2v}z}\left[A_{1b}
    \sin\left(\frac{M}{2v}z\right) + B_{1b}
    \cos\left(\frac{M}{2v}z\right)\right]
\end{eqnarray}
and similarly for $1 \leftrightarrow 2$. The boundary conditions
are $\phi_{1f}(0)$, $\phi_{1b}(L)$, $\phi_{2f}(0)$ and
$\phi_{2b}(L)$. Since the four fields are coupled by $R_0 =
B_{1f}B_{1b}^\star + B_{2f}B_{2b}^\star$ it is necessary to
consider all fields to determine the coefficients $A_{\mu i}$ and
$B_{\mu i}$. We start from Eqs.~(\ref{phimot}), we derive
expressions for $\phi_{jf}(0)$, $d\phi_{jf}(0)/dz$, $\phi_{jb}(L)$
and $d\phi_{jb}(0)/dz$, $j=1,2$ in terms of these coefficients and
use the relations (\ref{motion1}) to close this set of equations.

For $\sin(ML/2v), \phi_{1f}(0)$ and $\phi_{2f}(0) \neq 0$ we find
\begin{eqnarray}
    B_{1f} & = & \phi_{1f}(0)\nonumber \\
    B_{2f} & = & \phi_{2f}(0)\nonumber \\
    B_{1b} & = & -\frac{e^{-3i\varrho L/2v}M\left(\phi_{1b}(L)+
    \displaystyle\frac{\phi_{2b}(L)}{g\left(|B_{2f}|^2\right)}\right)}
    {\sin(ML/2v)\left[g(|B_{1f}|^2)+
\displaystyle\frac{4|B_{1f}|^2|B_{2f}|^2}{g(|B_{2f}|^2)}\right]}\nonumber \\
    B_{2b} & = & \frac{i2B_{1f}^\star B_{1b} B_{2f}-e^{-3i\varrho L/2v}
    \displaystyle\frac{M\phi_{2b}(L)}{\sin(ML/2v)}}{g(|B_{2f}|^2)}\nonumber
\\
    A_{1b} & = & \frac{\phi_{1b}(L)e^{-3i\varrho L/2v}-B_{1b}
    \cos\left(\frac{M}{2v}L\right)}{\sin\left(\frac{M}{2v}L\right)}\nonumber
\\
    A_{2b} & = & -B_{2b}/\tan(ML/2v)\nonumber \\
    A_{1f} & = & \frac{i}{M}\left[\varrho B_{1f} - 2\left(B_{1f}B_{1b}^\star
+
    B_{2f}B_{2b}^\star\right)B_{1b}\right]\nonumber \\
    A_{2f} & = & \frac{i}{M}\left[\varrho B_{2f} -
    2\left(B_{1f}B_{1b}^\star + B_{2f}B_{2b}^\star\right)B_{2b}\right]
\end{eqnarray}
where
\begin{equation}
    g(x) = i\left(\varrho - 2x\right) -
    M/\tan(ML/2v).
\end{equation}
These equations simplify considerably in the specific case
$\phi_{2b}(L)=0$ that we have analyzed in detail. Similar, but
simpler equations are easily derived if one of the input
fields is equal to zero or for $\sin(ML/2v) \neq 0$.

\end{document}